# Radiation Measurements Using Timepix3 with Silicon Sensor and Bare Chip in Proton Beams for FLASH Radiotherapy

**C. Oancea,**[a,1] **J. Šolc,**[b] **C. Granja,**[a] **E. Bodenstein,**[c,d] **F. Horst,**[c,d] **J. Pawelke,**[c,d] **J. Jakubek,**[a]

[a] *ADVACAM*
  *U Pergamenky 12, Prague 7, Czech Republic*
[b] *Czech Metrology Institute, Okruzni 31, 638 00 Brno, Czech Republic*
  *Okruzni 31, 638 00 Brno, Czech Republic*
[c] *OncoRay - National Center for Radiation Research in Oncology, Faculty of Medicine and University Hospital Carl Gustav Carus, Technische Universität Dresden*
  *Dresden, Germany*
[d] *Helmholtz-Zentrum Dresden-Rossendorf, Institute of Radiooncology - OncoRay*
  *Dresden, Germany*

[1]*E-mail*: cristina.oancea@advacam.cz

ABSTRACT: This study investigates the response of Timepix3 semiconductor pixel detectors in proton beams of varying intensities, with a focus on FLASH proton therapy. Using the Timepix3 application-specific integrated circuit (ASIC) chip, we measured the spatial and spectral characteristics of 220 MeV proton beams delivered in short pulses. The experimental setup involved Minipix readout electronics integrated with a Timepix3 chipboard in a flexible architecture, and an Advapix Timepix3 with a silicon sensor. Measurements were carried out with Timepix3 detectors equipped with experimental gallium arsenide (GaAs) and silicon (Si) sensors. We also investigated the response of a bare Timepix3 ASIC chip (without sensor). The detectors were placed within a waterproof holder attached to the positioning system of the IBA Blue water phantom, with additional measurements performed in air behind a 2 cm-thick polymethyl methacrylate (PMMA) phantom. The results demonstrated the capability of the Timepix3 detectors to measure time-over-threshold (ToT, deposited energy) and count rate (number of events) in both conventional and ultra-high-dose-rates (UHDR) proton beams. The bare ASIC chip configuration sustained up to a dose rate (DR) of 270 Gy/s, the maximum tested intensity, although it exhibited limited spatial resolution due to low detection efficiency. In contrast, Minipix Timepix3 with experimental GaAs sensors showed saturation at low DR ~5 Gy/s. Furthermore, the Advapix Timepix3 detector was used in both standard and customized configurations. In the standard configuration (Ikrum = 5), the detector showed saturation at DR ~5 Gy/s. But, in the customized configuration when the per-pixel discharging signal (called "Ikrum") was increased (Ikrum = 80), the detector demonstrated enhanced performance by reducing the duration of the ToT signal, allowing beam spot imaging up to DR= ~28 Gy/s in the plateau region of the Bragg curve. For such DR or higher, the frame acquisition time was reduced to the order of microseconds, meaning only a fraction of the pulse (with pulse lengths on the order of milliseconds) was captured.

KEYWORDS: Timepix3; FLASH proton beams; UHDpulse, proton radiotherapy, dose rate

**Contents**

**1. Introduction and goals**

Emerging cancer treatment techniques, such as FLASH proton therapy, present new challenges in dosimetry. Detectors and dosimetry protocols must be developed to handle ultra-short pulses of protons with the energy up to a few hundreds of MeV. Detectors based on the Timepix chip family have been applied in various particle therapy frameworks, including particle tracking [1], imaging [2], linear energy transfer (LET) spectrum measurements [3], mixed-field characterization [4-5] and treatment monitoring [6-7]. A key advantage of the Timepix3 detectors is their high detection efficiency and their ability to characterize radiation fields with spatial and temporal resolutions at the scale of a few micrometers and nanoseconds, respectively. In FLASH radiotherapy, both the delivered dose and the irradiation time are critical, and detectors capable of characterizing these parameters with precision are essential. As part of the UHDpulse project [8], a variety of detectors have been tested both in-beam and out-of-field. In this work, we evaluate a detector module in two configurations: the highly-integrated, miniaturized Minipix Timepix3 Flex and the high-performance, fast-readout Advapix Timepix3 (TPX3) rigid detector. For the Minipix TPX3 Flex detector, the chip's backing was fabricated from tissue-equivalent material to minimize any perturbations to the measured radiation field [9]. In contrast, the Advapix TPX3 detector is optimized for high-speed readout and designed with a rigid architecture [10]. The sensors used in this study were made from Si, GaAs, and a bare ASIC configuration. To directly assess the effects of radiation on the electronics, the ASIC chip was tested without a sensor. Each pixel in the Timepix3 system contains its own electronics, allowing particles interacting with the pixels to generate signals, primarily detected through the energy channel. The ASIC was used in its bare configuration, without any solid-state sensor attached. The UHDR beam required for FLASH irradiations can be sufficient to directly activate the readout pixels. The Timepix3 detector operates in frame mode, utilizing dual per-pixel signal electronics for simultaneous counting and energy measurements. This approach is particularly useful in high-intensity and UHDR radiation fields, where particles are registered as integrated pile-up events. This allows the particle flux and dose rates in high-intensity fields to be derived for both Si and GaAs sensors. The high granularity of the pixel detector, combined with its dual per-pixel signal acquisition and wide dynamic range in counting and energy, makes it well-suited for such measurements [11]. Each pixel can register event rates up to $10^4$ particles per second, which corresponds to a flux density of up to $10^8$ particles/cm$^2$/s assuming 32.5k pixels per cm$^2$ [11]. These values also account for charge sharing within the semiconductor pixels, which are operated at room temperature. The exact measurement range depends on the particle type; for instance, low-LET particles such as electrons and X-rays, which generate small tracks across only a few pixels, can be registered at higher fluxes. At higher particle rates (e.g., greater than $10^5$ particles/cm$^2$/s), event pile-up becomes significant, limiting the counting of individual events. However, the energy deposited by each particle is measured in an integrated mode, summing the contributions from all incoming particles during the frame acquisition time. This simultaneous registration of per-pixel energy deposition and event rates yields comprehensive spectrometric, dosimetric, and particle flux/intensity information. As a result, it enables detailed characterization of radiation fields even under high-intensity conditions. The goal of this work is to evaluate the performance of TPX3 detectors, both with and without sensors, in the context of FLASH proton therapy, thereby determining their potential for use in clinical applications of FLASH radiotherapy.

## 2. Instrumentation and measurements

### 2.1 Minipix TPX3 Flex and Rigid Advapix TPX3 detectors

The detectors used in this study were manufactured by ADVACAM a.s., Czech Republic, with the TPX3 chip developed under the Medipix Collaboration at CERN. The ASIC contains a matrix of $256 \times 256$ pixels, totaling 65,536 independent channels, where each pixel measures 55 µm. This provides an active sensor area of $14.08 \times 14.08$ mm$^2$. Figure 1 shows two TPX3 detector configurations. For this study, the sensors were fabricated from GaAs and Si, and one configuration utilized a bare ASIC to study direct radiation effects on the electronics. Figure 1a illustrates the Minipix TPX3 Flex detector configuration. The Si sensors had thicknesses of 500 µm, while the experimental GaAs sensors were 550 µm thick.

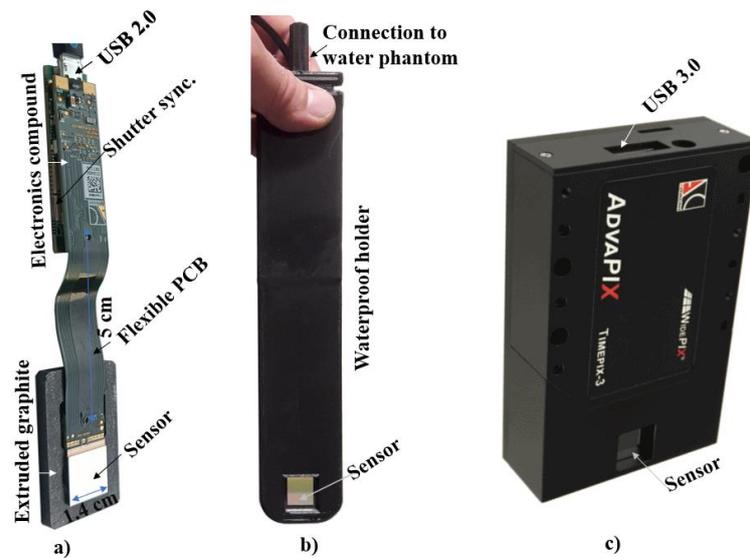

**Figure 1.** The TPX3 detector consists of a semiconductor radiation-sensitive Si sensor (500 µm thickness, 1.4 cm × 1.4 cm) bump-bonded to the ASIC TPX3 readout chip, featuring a $256 \times 256$ pixel array. a) The customized Minipix TPX3 Flex, which is free of metal holders, screws, and cooling elements. All metallic parts have been replaced by carbon and ABS plastic. The readout electronics are separated from the chip-sensor assembly using a 5 cm flexible connector attached to the printed circuit board (PCB) chipboard to minimize scattering and self-shielding. b) A 3D-printed holder designed to submerge the chip-sensor assembly in water. The assembly is enclosed in a silicon cover to make it waterproof. c) The rigid Advapix TPX3 detector.

Measurements were conducted to determine the most suitable detector configuration for primary beam characterization. These tests examined operational parameters such as detector mode, acquisition time, threshold settings, per-pixel saturation limits, and the impact of sensor material and thickness. Compared to the rigid Advapix TPX3 detector (Fig. 1c) with a Si sensor of 300 µm [10], the customized MiniPIX Flex version presents a characteristic design: the sensor is displaced 5 cm from the readout electronics using a flexible PCB. To further reduce external perturbation, all metal components from the case were replaced with plastic, and the sensor was mounted on an extruded graphite support, as shown in Figure 1b.

**2.2 Bare sensor Timepix3 detector**

These hybrid semiconductor pixel detectors can be also operated without the semiconductor sensor, bare ASIC chip [17-19]. The bare Minipix TPX3 detector was tested using two motherboard configurations which allow for using a positive and negative voltage. The results presented in this study were measured using the positive configuration.

**2.3 Detector operation and readout modes**

The Timepix3 detectors operate each pixel simultaneously in both time and energy modes or energy and counting mode. The time of arrival (ToA) can be determined with a resolution of 1.6 ns, while the ToT of the respective pixel can also be measured. For higher particle fluxes, the detector can be operated in frame mode using ToT and counting channels (Event + iToT), where it records the total/integrated per-pixel deposited energy (iToT, providing spectral and dosimetric information) and the number of hits in each pixel (events/counts, providing information about beam intensity). The detector's limitations are defined by the maximum iToT spectrometric signal that can be registered within a frame, which is approximately 16,300 iToT per pixel for this configuration. In this study, both detectors were operated in frame mode (Event + iToT) with a 4 keV threshold and a bias voltage as recommended during calibration (e.g., +200 V for the 300 µm and 500 µm Si sensors). Data were saved as individual files in ".txt" format. The threshold and pixel-by-pixel calibration were performed by ADVACAM using X-ray radiation sources of different energies [12].

**2.4 Experimental setup at the University Proton Therapy Dresden (UPTD)**

The experiments were performed at the University Proton Therapy Dresden, Germany. The proton therapy facility is equipped with a separate experimental room beside the patient treatment room. Our measurements were performed in the experimental room using a horizontal beamline, which delivers stationary proton pencil beams generated by an isochronous cyclotron (C230, Ion Beam Applications SA, Louvain-la-Neuve, Belgium). Proton beams with an energy of 220 MeV were delivered either with conventional dose rates or short UHDR (> 100 Gy/s) beam pulses. The uncollimated proton beam was directed perpendicularly to the central position of the front wall of an IBA Blue Phantom$^2$ (IBA Dosimetry, Schwarzenbruck, Germany) water phantom positioned at 116 cm downstream from the beam exit window. As determined from the Gaussian fit of dose profiles measured with a radiochromic film placed onto the front wall of the water phantom, the 220 MeV proton beam showed a full-width at half maximum spot size of 1.25 cm and 1.62 cm in horizontal and vertical direction, respectively. The dose rate (DR) in terms of absorbed dose-to-water at the reference point (Position B) was measured using a calibrated PTW ionization Semiflex chamber (PTW type 31010) placed inside the water phantom. The dose delivery to the phantom was monitored online by two ionization chambers (ICs): a beam monitor IC, integrated into the beam exit window, and an additional transmission IC (TM-IC, PTW type 7862) measuring beam monitor units (MU) and total ionization charge $Q_{TC}$, respectively. More details on the accelerator and beam characteristics of this setup can be found elsewhere [4].

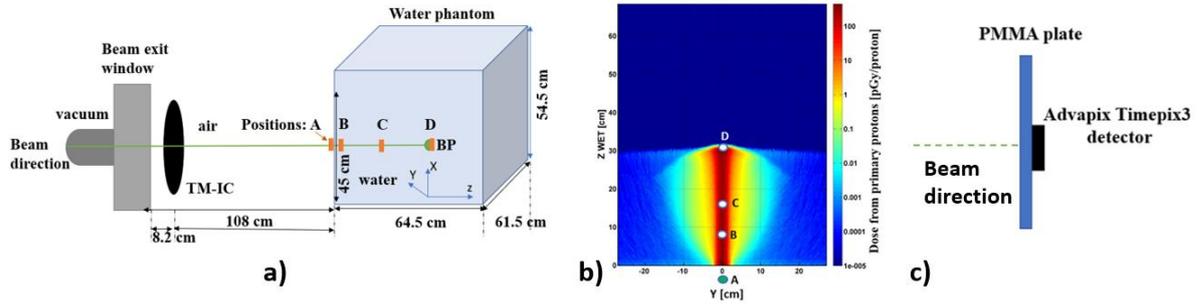

**Figure 2.** The experimental setup at the horizontal research beamline at University Proton Therapy Dresden. a) schematic diagram of the setup, from left to right: vacuum tube, beam exit window with integrated beam monitor chamber, transmission IC, air (~ 108 cm) and the water phantom (Blue Phantom²) also showing its coordinate system (X: lateral vertical, Y: lateral horizontal and Z: depth). b) Monte Carlo simulations of dose from primary protons in water phantom with identification of positions of interest where the detectors were placed [4, 13]. c) Schematic diagram showing the setup in air, beam direction, air (~ 50 cm), the PMMA plate of 2 cm thickness and a single detector Advapix TPX3 Si 300 µm thickness.

To achieve the FLASH dose rates, the accelerator was operated at high beam currents, reaching up to~100 nA at the beam exit window, and irradiation was performed in short pulses. The TPX3 detectors were placed in the water phantom at different positions. Position A: At the entrance to the water phantom in air, Position B (reference point where DR was measured): 2 cm behind the water phantom wall, corresponding to a 5 cm water equivalent thickness (WET), Position C: In the plateau region of the Bragg curve, Position D: In the Bragg peak region. The exact detector positions in terms of WET are shown in Figure 2b. The aim was to measure the integrated per-pixel ToT and the number of events reaching the detector sensor. The experiment was conducted using three detectors: a Minipix TPX3 Flex with an ASIC without a sensor, with a Si sensor, and with a GaAs sensor. In another experimental setup, as shown in Figure 2c, the rigid Advapix TPX3 detector was used to image the beam and measure both ToT and count rate in two configurations: (i) standard per-pixel energy calibration with a threshold of 3 keV and Ikrum value of 5, and (ii) customized calibration with a threshold of 4 keV and Ikrum value of 80. In the customized configuration the per-pixel discharging signal was increased from the standard value of 5 to 80 to explore the impact of increased discharging signal on detector performance, particularly in terms of decreasing the per-pixel ToT to avoid saturation and measure under higher intensity conditions [14].

## 3. Results and Discussion

### 3.1 Bare Timepix3 ASIC

The detector with a bare ASIC chip was positioned directly in the proton beam, with the chip oriented perpendicular to the beam axis. Proton beams with a nominal energy of 220 MeV and gradually increasing DR within each pulse were delivered to evaluate the detector settings, operational limits, and saturation thresholds. Figure 3 presents a 2D visualization of (a) per-pixel radiation signal and (b) per-pixel event rate, measured using the Minipix TPX3 Flex with an ASIC without a sensor at the Bragg peak region, corresponding to the position D from Figure 2. Proton beams with pulse lengths of 10 ms and DR at the reference point of approximately ~30 Gy/s (Figure 2 left) and ~270 Gy/s (Figure 2 right) were delivered. The detection channel responsible for registering the radiation signal (iToT) was found to be more sensitive than the channel measuring the event rate, which corresponds to the number of pixel hits, as visualized in Figure

3. The results indicate that the bare TPX3 detector could operate effectively up to ~270 Gy per pulse (pulse length of 10 ms) in the plateau region at a beam current of ~97 nA.

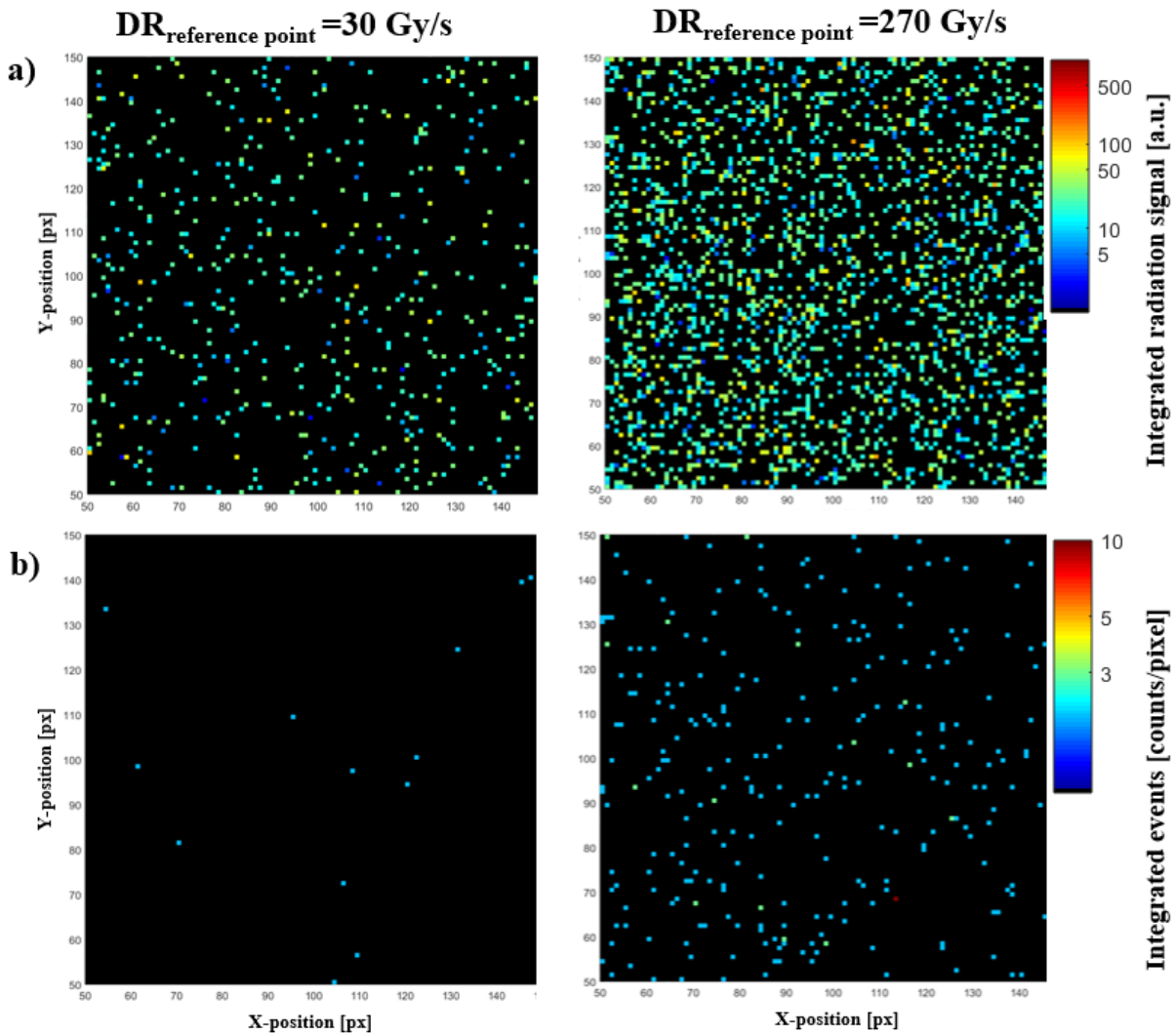

**Figure 3.** Detection and visualization by the bare TPX3 ASIC chip (without sensor) of the primary radiation field inside the water phantom by a primary 220 MeV proton beam at low-DR (left) and UHDR (right). Data measured for a pulse duration of 10 ms with the detector readout operated in frame mode showing the per-pixel a) energy and b) event-counting response. The detector was placed at Position D distal to the Bragg peak region. The per-pixel radiation signal and per-pixel event counts are displayed by the colour logarithmic scale. A selected area of the detector pixel matrix is displayed (100 × 100 pixels = 5.5 mm × 5.5 mm = 30 mm$^2$).

Figure 4 shows the evaluated signal of a bare Timepix3 ASIC to delivered pulse dose rate. It exhibits a linear response in the plateau region of the Bragg curve, with $R^2 = 0.9866$ at position C and $R^2 = 0.9816$ at position D. Figure 4b further illustrates the linearity of the event counting rate channel for various DR at two positions along the Bragg curve. The detector remained functional at dose rates up to ~270 Gy/s at the reference point. However, additional measurements are necessary to investigate whether the event count rate is directly proportional to physical quantities such as particle flux. In the negative configuration, the same ASIC without a sensor was found to be more sensitive to radiation, exhibiting saturation and malfunctioning more frequently compared to the positive configuration. Though, the linearity of the signal was improved in the positive configuration. Similar tests were conducted with a Minipix TPX3 Flex equipped with a 500 µm Si sensor and experimental GaAs sensors. The results for the Si sensor demonstrated that the detector could be operated in frame mode up to UHDR, but saturation occurred. Additional measurements were performed using detectors with GaAs sensors in primary proton beams. The measured data from these detectors exhibited saturation at a lower threshold, before reaching UHDR conditions. This behavior is primarily due to the high atomic number (Z) and density of the GaAs material, which results in significantly greater energy loss and higher deposited energy in the semiconductor detector. GaAs material was chosen for its superior radiation hardness compared to Si.

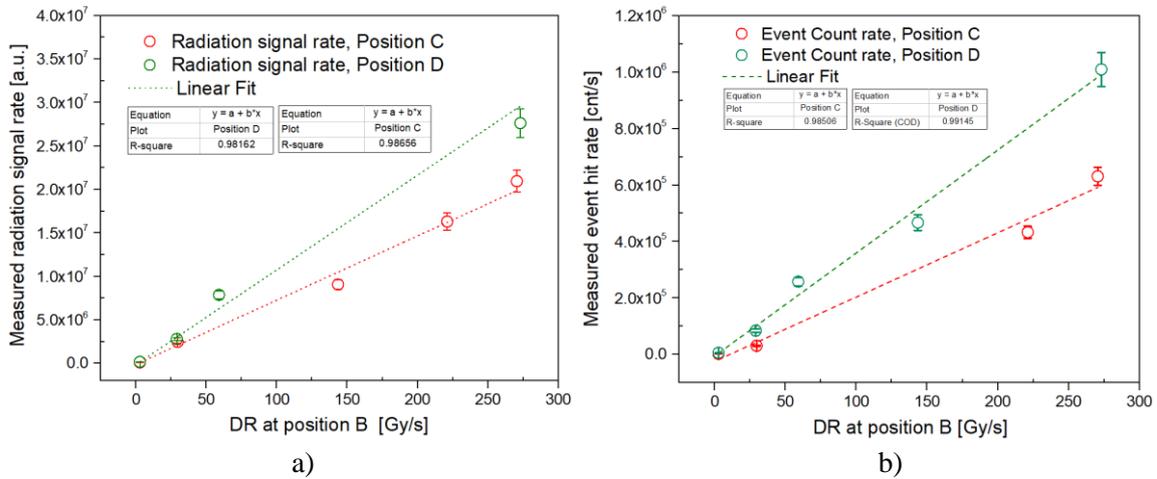

**Figure 4.** a) Measured radiation signal (summed over an area of 160×245 pixels) using the energy detection channel by the Minipix TPX3 detector with a bare ASIC chip (Y-axes) and the delivered DR at the reference point (X-axes) at two positions along the Bragg curve: position C (entrance region of the Bragg curve) and position D (distal to the Bragg peak region) from Fig 2c. b) Registered per-pixel event rate for the same two positions along the Bragg curve inside the water phantom (position C and D) measured with the bare Minipix TPX3 Flex detector (positive settings). Parameters of the linear fit are shown.

### 3.2 Advapix Timepix3 ASIC with a Si sensor

Figure 5 shows the 2D visualization of the per-pixel integrated iToT measured with the Advapix TPX3 Si sensor (300 µm) in air, positioned in the proton beam behind a 2 cm-thick PMMA plate. Data are the averaged values of various frames of 500 µs acquisition time. The comparison between the two detector configurations shows the effect of increased per-pixel discharging signal (Ikrum = 80) on the detector ability to measure the ToT signal and avoid saturation. By increasing the Ikrum values the ToT signal is shortened, which results in the ability to measure more particles. A clear beam spot profile is observed at DR = 0.38 Gy/s in both configurations. However, when using the standard configuration (Ikrum = 5), saturation is already evident at DR = 5.5 Gy/s, resulting in distortion of the measured beam profile due to the limited dynamic range. At lower Ikrum values, the ToT signal is saturated, leading to further distortion of the beam profile at higher DRs. In contrast, the customized configuration (Ikrum = 80) increases the dynamic range, allowing for imaging of beam profile at higher DR without saturation, making it more suitable for UHDR beam measurements. The detector in this customized configuration could be used at DR reaching the FLASH regime at the Bragg peak. In Figure 6a, where a DR of ~28.4

Gy/s was delivered, the detector successfully imaged the beam spot profile with a lower acquisition time of 50 µs, capturing only a small portion of the 3 ms beam pulse.

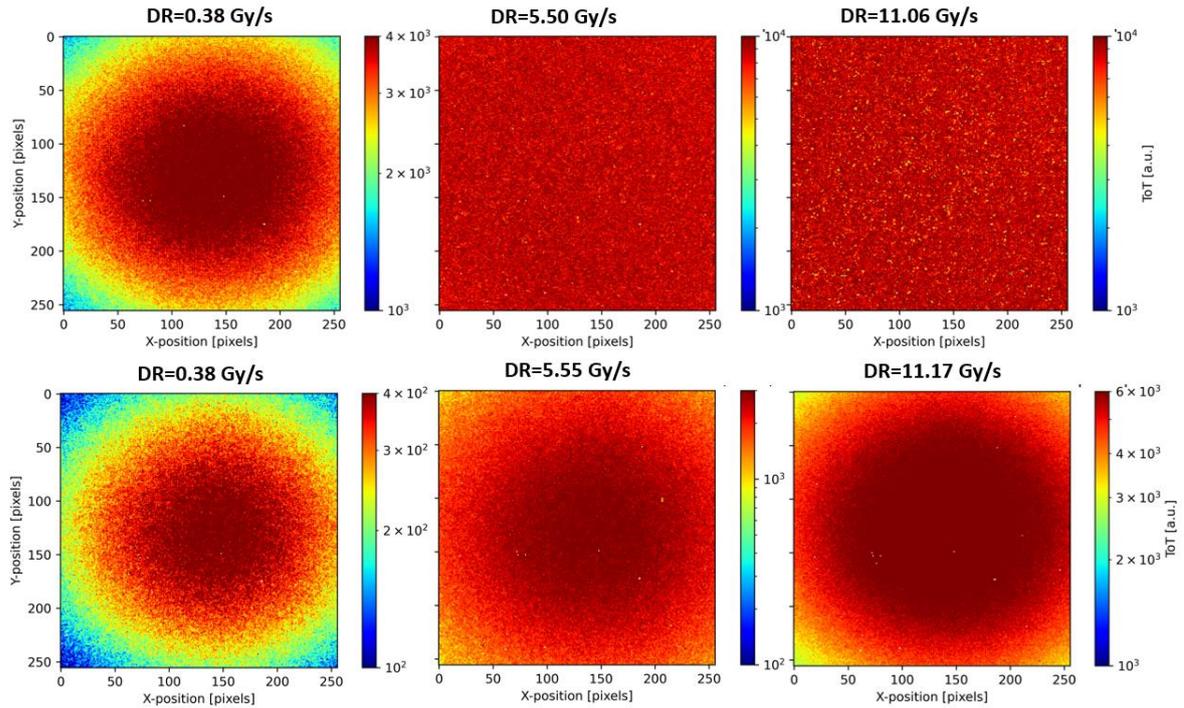

**Figure 5.** 2D visualization of per-pixel integrated ToT measured with the Advapix Timepix3 Si sensor of 300 µm in air behind 2 cm-thick PMMA. The top row illustrates the iToT measurements acquired in frame mode (averaged over 500 µs frames duration) obtained using the standard configuration of the detector with Ikrum set to 5, across different delivered dose rates (DR = 0.38 Gy/s, 5.50 Gy/s, and 11.06 Gy/s). The bottom row displays the iToT measurements for the same DR, but using a customized detector configuration with Ikrum set to 80. The color logarithmic scale represents the iToT values/bins for a frame duration of 500 µs.

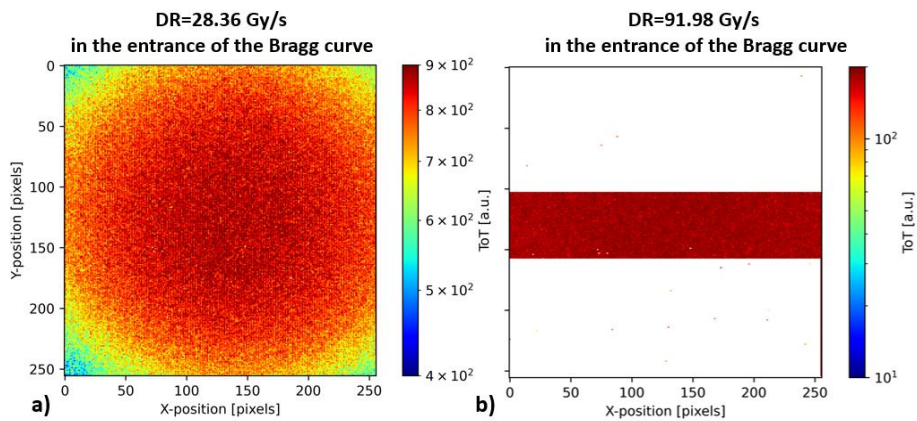

**Figure 6.** Same as Figure 5 (Bottom): iToT measured with a customized configuration at Ikrum = 80 for higher dose rates, reaching UHDR at the Bragg peak region. In a), the entire sensor area was used, and the frame acquisition time was reduced to 50 µs. In b), a mask was applied over the sensor area, disabling a portion of the sensor to limit the active area and reduce the data rate. The active area consisted of 50 × 256 pixels and the acquisition time was decreased to 10 µs.

The frame acquisition time of the detector defines the saturation threshold and limits its response. As the DR increases, the acquisition time must be shortened to prevent sensor saturation. When converting ToT values to deposited energy or dose, this short frame acquisition time should be considered and corrected [15]. As the DR was increased to ~92 Gy/s, further reduction in acquisition time to 10 µs was necessary, and a mask was applied over the sensor to limit the active area and reduce the data rate. Rows of the sensor were masked to increase the detector speed, leaving an active area of 50 × 256 pixels. However, at this UHDR, the detector began to experience performance issues, and the short acquisition time led to underestimation of the ToT values, as the charge from particles was only partially collected [16].

## 4. Discussion and conclusions

In this study, the performance of Timepix3 detectors, both with and without sensors, was evaluated for their ability to characterize proton beams at UHDR with potential application in FLASH proton therapy. The bare Timepix3 detector with positive settings effectively operated at maximum dose levels of ~270 Gy/pulse without experiencing saturation, exhibiting a linear response. The GaAs sensor showed saturation at DR above ~5.5 Gy/s. The Advapix TPX3 with a Si sensor demonstrated improved performance under a customized configuration with an increased Ikrum value (80), enabling measurements at higher DR while mitigating saturation effects. However, further studies are necessary to optimize detector performance and improve the accuracy of converting measured radiation signals into dosimetric quantities.

### Acknowledgments:

We would like to express our gratitude to Jiri Pivec for his valuable contribution to the experiment and data collection.